\documentclass{jfm}
\usepackage{graphicx}
\usepackage{bm}
\usepackage{psfrag,amssymb}
\usepackage{amsmath}

\def\etal{\mbox{\it et al.\ }}
\begin{document}
 \title{Viscous lock-exchange in rectangular channels}

  \author[J. Martin \etal]{
	J\ls.\ns M\ls A\ls R\ls T\ls I\ls N,
	N\ls.\ns R\ls A\ls K\ls O\ls T\ls O\ls M\ls A\ls L\ls A\ls L\ls A\ls,
	L\ls.\ns T\ls A\ls L\ls O\ls N  \and
	D\ls.\ns S\ls A\ls L\ls I\ls N}
  \affiliation{Univ. Pierre et Marie Curie-Paris6, Univ. Paris-Sud, CNRS.\\
   Lab FAST, Bat. 502, Rue du Belvedere, Campus Univ., Orsay, F-91405, France.}
  \date{\today}

 \maketitle


 \begin{abstract}
In a viscous lock-exchange gravity current, which describes the
reciprocal exchange of two fluids of different densities in a
horizontal channel, the front between two Newtonian fluids
spreads as the square root of time. The resulting diffusion coefficient
reflects the competition between the buoyancy driving effect and the viscous damping, and depends on the geometry of the channel. This lock-exchange diffusion coefficient
has already been computed for a porous medium, a $2D$ Stokes flow between
two parallel horizontal boundaries separated by a vertical height, $H$, and, recently, for a cylindrical tube. In the present paper, we calculate it, analytically, for a rectangular channel (horizontal thickness $b$, vertical height $H$) of any aspect ratio ($H/b$) and compare our results with experiments in horizontal rectangular channels for a wide range of aspect ratios ($1/10-10$).
We also discuss the $2D$ Stokes-Darcy model for flows in Hele-Shaw cells and show that it leads to a rather good approximation, when an appropriate Brinkman correction is used.
 \end{abstract}
\section{Introduction}

\begin{figure}
\begin{center}
 \includegraphics[width=14cm]{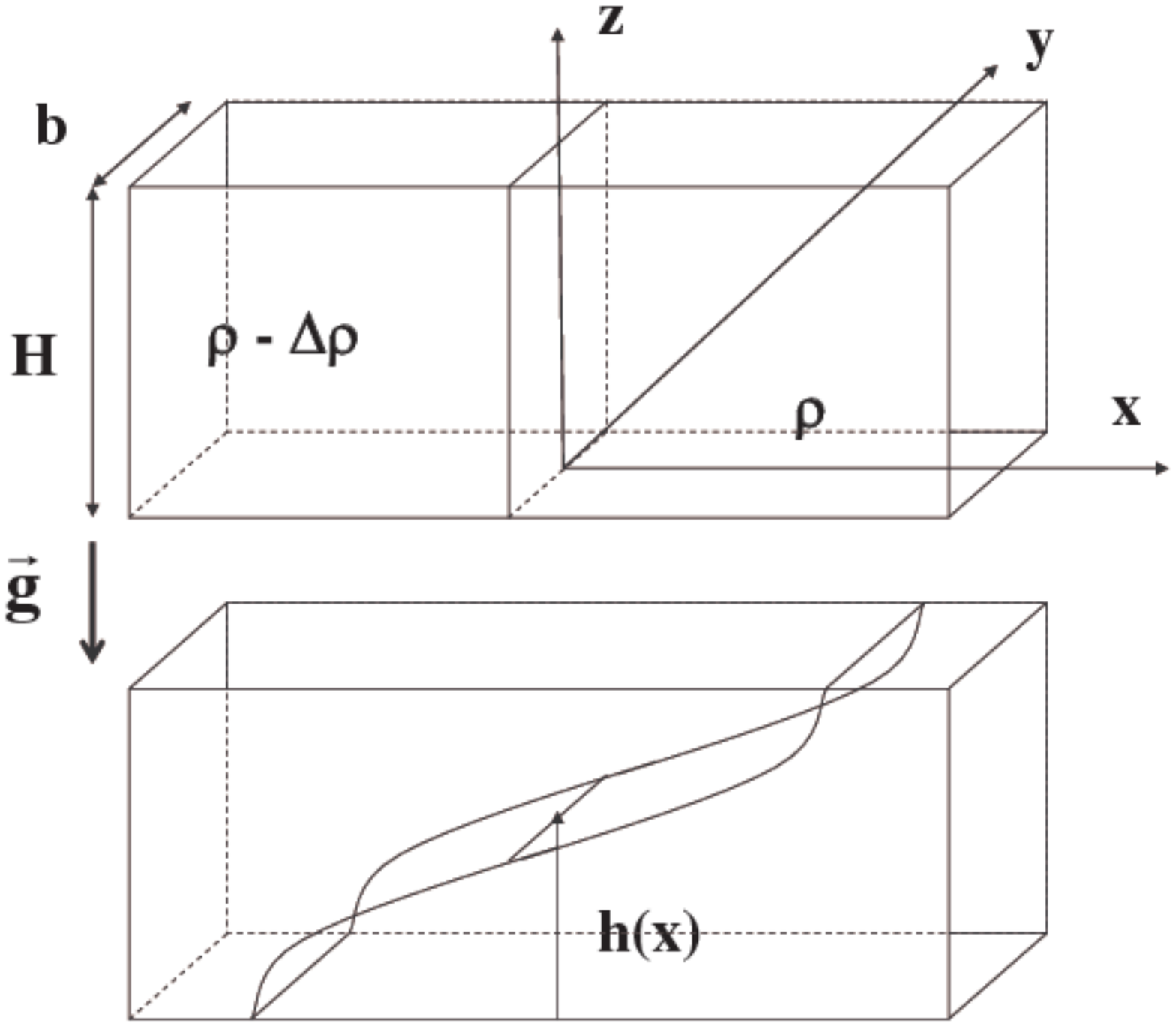}
 \caption{Sketch of the rectangular cell of height $H$ in the gravity direction and width $b$.
Top: Initial configuration where the heavy fluid of density, $\rho$, is separated
from the light fluid of density, $\rho-\Delta\rho$, by a vertical gate (plane $y-z$).
Bottom: After removing the gate, buoyancy differences cause the denser fluid to
flow in one direction along the bottom of vessel, while the lighter one flows
in the opposite direction at the top of the vessel. The goal is to determine
the space and time dependencies of the pseudo-interface, $h(x,t)$.}
           \label{schema}
\end{center}
\end{figure}

The lock-exchange configuration refers to the release, under gravity,
of the interface between two fluids of different densities, confined in the section of a horizontal channel. This physical process has prompted renewed interest, as a part of the carbon dioxide
sequestration issues (\cite{neufeld09}).
 The top of Fig. \ref{schema} shows the initial lock-exchange
situation of a so-called full-depth release. The two fluids, initially separated by a vertical barrier (the lock gate), fill the whole section of the tank.
When the gate is withdrawn (bottom of Fig. \ref{schema}), buoyancy drives the denser fluid along the bottom wall, while the lighter one flows in the opposite direction at the top of the channel. The so-called lock-exchange results in the elongation of the interface between the two fluids along the horizontal direction.
Different regimes have been reported for the velocity and shape of the elongating interface.
The slumping phase refers to the initial regime where inertia dominates over viscous forces,
which typically applies for the case of salted and fresh water in a tank.
In this regime, \cite{benjamin68}, and more recently \cite{shin04} showed that
in the presence of a small density contrast (i.e. in the Boussinesq approximation $\Delta\rho << \rho$), the two opposite currents traveled at the same constant velocity.
When the Boussinesq approximation does not apply, \cite{lowe05}, \cite{birman05}, \cite{cantero07}
and \cite{bonometti08} showed that the two opposite fronts did travel at constant, but with different velocities. However this interface elongation, proportional to the time,
is slowed down at later stages, in the viscous phase, where dissipation prevails over inertia.
In the latter regime, the interface elongates as $t^\alpha$, where the exponent $\alpha$, smaller than unity, may take different values depending on the geometry and
the confinement of the flow (\cite{didden82, huppert82, gratton90, cantero07, takagi07, hallez09}).
In porous media, \cite{bear88} and \cite{huppert95} predicted an interface
spreading proportionally to the square root of time that \cite{seon07} observed in a horizontal cylindrical tube. Such a spreading can be quantified with
a diffusion coefficient, which reflects the balance between the buoyancy driving and the viscous damping. This coefficient, which depends on the nature and the geometry
of the flow, has been computed for a porous medium by \cite{huppert95},
for a $2D$ Stokes flow between two parallel horizontal boundaries separated
by a vertical height, $H$, by \cite{hinch07} and \cite{taghavi09},
and for a cylindrical tube by \cite{seon07}. However, to our knowledge,
such a diffusion coefficient has not been derived for a rectangular channel
(horizontal thickness $b$, vertical height $H$, Fig. \ref{schema}), for which
one expects to recover the porous medium regime for $b \ll H$, and, possibly,
the Stokes flow regime for $b \gg H$.
In order to gather the limiting cases in the same paper, we first recall the
results for porous media and $2D$ Stokes flows, together with the tube case,
for the sake of comparison. Then we compute, for a rectangular channel of aspect ratio, $H/b$, the dependence of the interface $h(x,t)$ and the corresponding viscous lock-exchange diffusion coefficient. We also test the so-called Stokes-Darcy $2D$ model to this lock-exchange configuration. Finally, we test and validate our theoretical results
with experiments in horizontal rectangular channels for a wide range of aspect ratios ($1/10-10$).

\section{Lock-exchange in different geometries }

Let us first recall the basic hypotheses on the viscous gravity currents,
common to the different geometries, used for instance by \cite{huppert95} or \cite{hinch07}).
As sketched in Fig. \ref{schema}, the interface between the two
fluids is assumed independent on the $y$ direction. Its distance from the bottom boundary of the vessel is denoted $h(x,t)$.
 This interface can be a pseudo-interface between two miscible fluids
for which molecular diffusion can be neglected or between two immiscible fluids, provided that the interfacial tension can be neglected.
 The flow is assumed to be quasi-parallel to the horizontal $x$ axis. This is a key hypothesis.
Neglecting accordingly the vertical component of the fluid velocity implies that the vertical pressure gradient follows the
hydrostatic variations: $\partial p/ \partial z = - \rho g $.
This hypothesis is violated at short times, immediately after the opening of the gate, but should become valid at later stages,
as soon as the interface has slumped over a distance larger than $H$, thus
ensuring a small enough local slope $\partial h/\partial x$.
Then, the pressures, $P_+$ and $P_-$, in the lower layer, $0< z < h(x,t)$,
and in the upper one, $h(x,t)< z <H$, respectively, write
\begin{equation} \label{pressure}
P_{+}=p(x,t)- \rho_{+} g z \;\; ; \; P_{-}=P_{+}+\Delta\rho g
(z-h(x,t))
\end{equation}
where $p(x,t)$ denotes the pressure at the lower wall, $z=0$. The difference between the horizontal pressure gradients in the two fluids is therefore linked to the interface slope by:
\begin{equation} \label{pressuregrad}
\frac{\partial P_{+}}{\partial x} - \frac{\partial P_{-}}{\partial
x}=\Delta\rho g \frac{\partial h}{\partial x}
\end{equation}
The time evolution of the interface, $h(x,t)$, is governed by the
mass conservation of each fluid (see Fig. \ref{schema} for notations).
For instance, for the heavier bottom layer we have:
\begin{equation} \label{conservation1}
\frac{\partial h}{\partial t}+\frac{\partial q}{\partial x}=0
\end{equation}
where $q(x,t)=q_{+}(x,t)$ is the horizontal flux ($m^{2}/s$) of the denser fluid at the location $x$:
\begin{equation} \label{flux}
q(x,t)=q_{+}(x,t)=\int_{0}^{h} \frac{1}{b}\int u_x(x,y,z,t)dy dz
\end{equation}
with $u_x(x,y,z,t)$ the $x-$velocity component and $b$ the spanwise length
(the $y$ integration is along this spanwise length). Moreover, in our configuration of uniform section along the horizontal axis $x$, $u_x(x,y,z,t)$ must also satisfy the no net flux condition:
\begin{equation} \label{netflux}
q_{+}(x,t)+q_{-}(x,t)=\int_{0}^{H} \frac{1}{b}\int u_x(x,y,z,t)dy dz=0
\end{equation}
We will see in the following that, in the viscous regime of interest,
the horizontal velocity component $u_x$, solution of either a Darcy or a Stokes equation, is proportional to the pressure gradient in each fluid layer.
Such solutions, combined with eq. (\ref{pressuregrad}), eq. (\ref{flux})
and eq. (\ref{netflux}) allow then to eliminate the pressure gradients and
to derive an expression of the flux $q$, of the form:
\begin{equation} \label{diffusiveflux}
q =- D f(\frac{h}{H}) \frac{\partial h}{\partial x}
\end{equation}
where $D$ writes
\begin{equation} \label{d}
 D=\tau \frac{\Delta \rho g }{\eta}
\end{equation}
and where the constant $\tau$ (scaling with a volume) and the function $f$ depend on the geometry and the flow equation and $\eta$ is the dynamic viscosity. Using the expression (\ref{diffusiveflux}) for the flux,
eq. (\ref{conservation1}) admits a self-similar solution
$h(\zeta)=H\, \psi(\zeta)$ with the similarity variable $\zeta=x/\sqrt{D t}$, which obeys:
\begin{equation} \label{similarity}
-\zeta \frac{d \psi}{d \zeta}=2 \, \frac{d}{d \zeta}(f(\psi) \frac{d \psi}{d \zeta})
\end{equation}
This equation may alternatively be rewritten, in terms of $\zeta (\psi)$:
 \begin{equation} \label{similaritybis}
\zeta (\frac{d \zeta}{d \psi})^{2}-2 \, f
\frac{d^{2}\zeta}{d \psi^{2}}+2\frac{d f}{d\psi}(\frac{d\zeta}{d \psi})=0
\end{equation}
The solution of the above equations can be found analytically or numerically,
depending on the complexity of the normalized flux function $f(\psi)$.
In the following, we will first recall the case of porous media,
treated by \cite{huppert95} and the $2D$ Stokes flow, addressed by \cite{hinch07} (unpublished)
and \cite{taghavi09}. We note that the latter paper included the effects of
the rheological properties of the fluids. However, in order to focus on the geometrical aspects, we will assume in the following that both fluids are Newtonian and have the same viscosity.

\subsection{Lock-exchange in porous media}

\begin{figure}
\begin{center}
 \includegraphics[width=14cm]{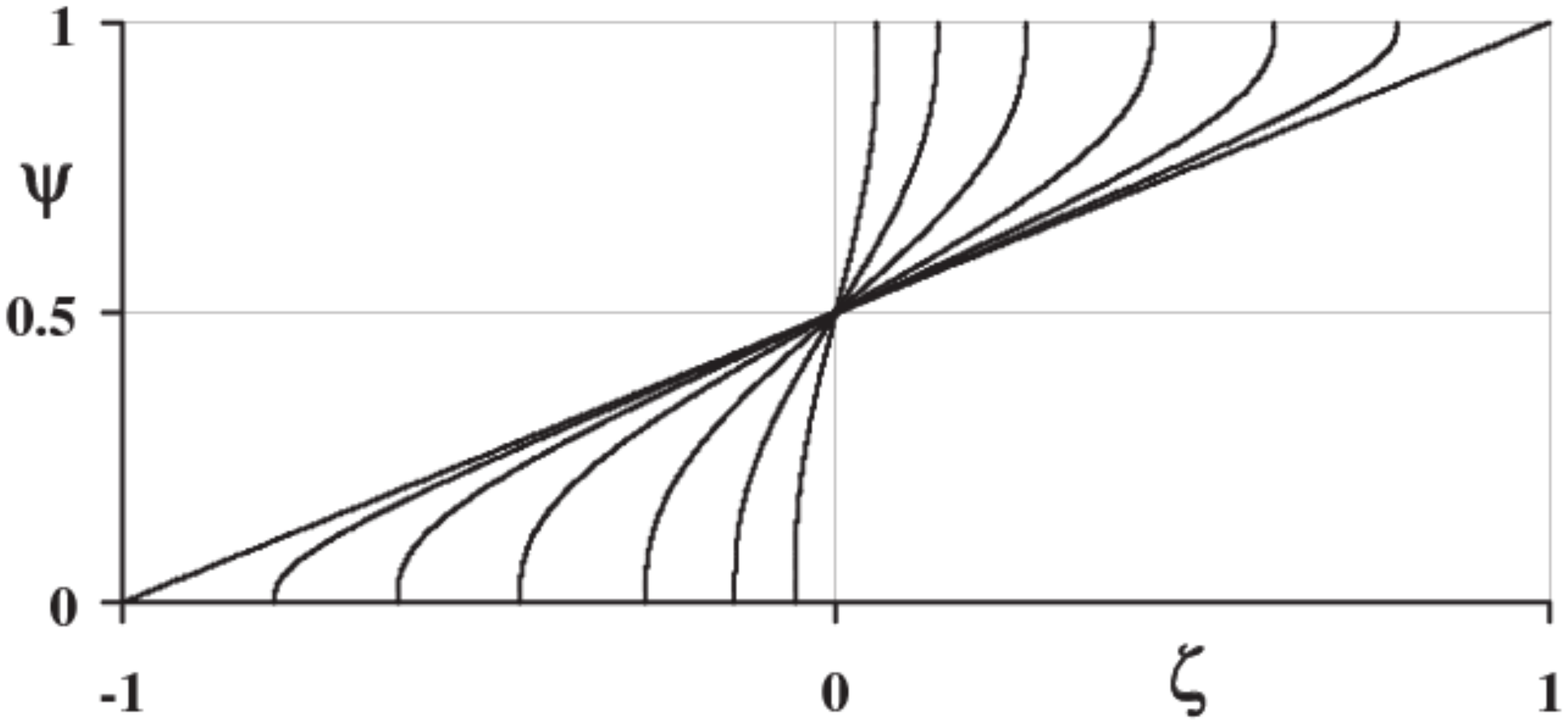}
 \caption{Lock-exchange interface $\psi(\zeta)$ between the two fluids for rectangular
            cells of different aspect ratios $\Gamma=H/b$.
The straight line of slope 0.5 corresponds to the Darcy porous media limit
(eq. (\ref{profilDarcy})). The other curves  correspond to the case of a rectangular
cross-section channel (subsection \ref{rectangular channel}).
From the bottom left to right,  $\Gamma=10, 4, 2, 1, 0.5$ and $0.2$.}
\label{GCprofil}
 \end{center}
\end{figure}

For a homogeneous layer of porous medium of permeability $\kappa$ (see for
instance \cite{huppert95}),
 the flow in each fluid is given by Darcy's law which relates the velocity in
each phase to the local pressure gradient :
\begin{equation} \label{darcy}
	u_{x \pm}=-\frac{\kappa}{\eta}\frac{\partial P_{\pm}}{\partial x}
\end{equation}
At a given location $x$, the velocity is then uniform in each layer, and the
no net flux condition (eq. (\ref{netflux})) simply writes:
$h u_{x +}+(H-h)u_{x -}=0$. The latter equation, combined with eq. (\ref{darcy})
and eq. (\ref{pressuregrad}), leads to  eq. (\ref{diffusiveflux}), and thus
(combined with eq. (\ref{conservation1})) to eq. (\ref{similarity})
with a diffusion coefficient and a flux function:

\begin{eqnarray}
  & & D_{PM}=\kappa\, H\frac{\Delta \rho g }{\eta}  \label{dDarcy}\\
  & & f_{PM}(\psi)=\psi (1-\psi) \label{fluxDarcy}
\end{eqnarray}
The solution of eq. (\ref{similarity}), in the similarity variable $\zeta=x/\sqrt{D_{PM} t}$,
is then a linear profile (\cite{huppert95}):
\begin{equation}\label{profilDarcy}
    \psi=h(x,t)/H=(1+\zeta)/2
\end{equation}
The so-obtained front profile in homogeneous porous media is displayed in
Fig. \ref{GCprofil} (straight line) together with the ones for rectangular cells (referred to in subsection \ref{rectangular channel}).
The leading ($\psi=0$, $\zeta=-1$) and trailing ($\psi=1$, $\zeta=1$) edges
spread as $\sqrt{D_{PM} t}$. Therefore, the lock-exchange diffusion coefficient
for porous media is $D_{PM}$. It should be noticed that \cite{bear88} reported a numerical integration of eq. (\ref{similarity}) indicating that the gravity current spreads as the square root of time.
Note that the corresponding result for a Hele-Shaw cell, that is two parallel plates of height,
$H$, separated  by a tiny gap $b$ ($b\ll H$), is obtained
using the permeability $\kappa=b^{2}/12$:

\begin{equation}\label{diffdarcy}
    D_{HS}=\frac{b ^{2}H \Delta \rho g }{12 \eta}
\end{equation}

\subsection{Lock-exchange for a $2D$ Stokes flow between two horizontal boundaries}

For a $2D$ Stokes flow between two horizontal parallel boundaries,
separated by a height $H$ in the plane ($z-x$) (assuming invariance along the
$y-$direction), the flow in each fluid is given by the Stokes equation:
\begin{equation} \label{stokes}
\eta \; \nabla^{2} u_{x \pm}(x, z)=\frac{\partial P_{\pm}}{\partial x}
\end{equation}
At a given location $x$, the velocity profile consists of two parabola
profiles matching the no slip boundary conditions at the bottom and the top boundaries
($u_{x +}(x, 0)=u_{x -}(x, H)=0$) and the continuity of the velocity
($u_{x -}(x, h)=u_{x +}(x, h)$) and of the shear stress
($\eta \; \partial u_{x -}(x, h)/\partial z= \eta \; \partial
u_{x +}(x, h)/\partial z$) at the interface, $z=h$.

\noindent Using the no net flux condition (eq. (\ref{netflux}))  and eq. (\ref{pressuregrad}),
 we obtain eq. (\ref{diffusiveflux}), which enables to rewrite eq. (\ref{conservation1}),
in the form of eq. (\ref{similarity}) or eq. (\ref{similaritybis}), with:

\begin{eqnarray}
  & & D=\frac{H^3 \Delta \rho g}{3 \eta}  \label{d2D}\\
  & & f(\psi)=\psi^{3} (1-\psi)^{3} \label{flux2D}
\end{eqnarray}
Note that the polynomial development of the solution of eq. (\ref{similaritybis})
around $\psi=0$ gives: $\zeta=-\zeta_0 +2\psi^{3}/(3 \,\zeta_0)$.
Thus, the location, $\zeta(\psi=0)=-\zeta_0$, of the leading edge of the interface is indeed constant in the similarity variable. Moreover, the development shows that
the slope of the interface is vertical at the bottom wall ($\psi=0$).
This is also the case at the upper wall ($\psi=1$), as
the problem is symmetric with respect to the centre of the cell.
We note that in the presence of such a vertical slope, our (horizontal)
quasi-parallel flow assumption falls locally,
but it is still valid upstream and downstream, where the slope of the interface remains small.
The solution of eq. (\ref{similaritybis}) can be found numerically using a shooting method similar to the one used by \cite{hinch07} and \cite{taghavi09}.
It was computed starting the integration of eq. (\ref{similaritybis})
from $(\psi=0.5,\zeta=0)$ and matching the asymptotic development in the vicinity of $\psi=0$.
From the so-obtained solution, one can deduce the spreading diffusion coefficient between the leading edge ($h=0$, $-\zeta_0=-0.1607$) and the trailing edge ($h=H$, $\zeta_0$)
of the front, from $[0.5 \,(x(h=0)-x(h=H))]^2=D \, t$,
which gives $D_{2D}=D\,\zeta_0^2$, so that:
\begin{equation}\label{D2Dstokes}
    D_{2D}=0.0086\, H^{3}\,\frac{\Delta \rho g }{\eta}
\end{equation}
This result is in agreement with the one found by \cite{hinch07}. \cite{taghavi09} provide five $\psi(\zeta)$ plots in their Fig. 9, corresponding to different viscosity ratios and including our case. From that figure we may obtain a value of their similarity variable,
 $\eta_0\sim 0.09$, which is consistent with our finding $\zeta_0=0.1607$,
when taking into account their definition of the similarity variable, $\zeta=\eta \,\sqrt{3}$.

For completeness, $D_{2D}$ may be compared to the result for a cylindrical tube
of diameter $d$ (\cite{seon07}):
\begin{equation}\label{Dtube}
    D_{T}= 0.0054 \,d^{3}\,\frac{\Delta \rho g }{\eta}
\end{equation}
The above expression is indeed very close to the $2D$ result, with $d$ playing the role of $H$.

\subsection{Lock-exchange in a rectangular cross-section channel}
\label{rectangular channel}

This article aims to extend the computation of the lock-exchange diffusion
coefficient to rectangular cells of arbitrary cross-sections
$H\times b$ (see Fig. \ref{schema}).
In the following, the cross-section aspect ratio  will be denoted $\Gamma= H/b$.
As in the previous section, a quasi-parallel flow approximation is
assumed (\emph{i.e.} small interface slope) which leads to eq. (\ref{pressuregrad})
for the pressure gradient. We will also assume the invariance of the interface
location along the gap direction $y$. This requires that the deformation of the interface,
induced by the flow profile along the direction $y$, relaxes much more quickly
than the gradient along $x$.
The flow in each fluid obeys a $3D$ Stokes equation:

\begin{equation}\label{Stokes3D}
    \eta \; \nabla^{2} u_{x \pm}(x,y,z)=\frac{\partial P_{\pm}}{\partial x}
\end{equation}
In order to solve this equation, we follow the series decomposition in Fourier modes of the velocity field used by \cite{gondret97}. This paper addressed the issue of the
parallel flow of two fluids of different viscosities in a rectangular cell. This issue is very closed to ours, as it requires to solve the Poisson equation (eq. (\ref{Stokes3D})),
but with different viscosities and the same pressure gradient for both fluids in \cite{gondret97}.
The method used was to split the velocity into two terms,
$u_{x \pm}(x, y, z)=u_{x \pm}^{*}(x, y)+u_{x \pm}^{**}(x, y, z)$.
Here, the first term,
$u_{x \pm}^{*}(x, y)=
\frac{b^2}{8 \eta} \frac{\partial P_{\pm}}{\partial x} [1-(\frac{2 y}{b})^2]$
is the Poiseuille-like unperturbed velocity far away from the interface.
The second term satisfies the Laplace equation,
$\nabla^{2}u_{x \pm}^{**}(x, y, z)=0$ and
vanishes far away from the interface. Its expression in terms of a sum of Fourier modes leads to a velocity profile of the form:

\begin{multline}
u_{x \pm}(x, y, z)=\frac{b^2}{8 \eta}
\frac{\partial P_{\pm}}{\partial x} {\left\{ 1-(\frac{2 y}{b})^2 \right.} \\
\left.{+\sum_{n = 1}^{\infty}
\frac{32 (-1)^n \,(a_{\pm n} \, e^{(2n-1)\frac{\pi (z-H/2)}{b}}
+b_{\pm n} \, e^{-(2n-1)\frac{\pi (z-H/2)}{b}})}
{{\left (2\,n-1 \right)}^3 \, {\pi }^3} \cos[(2n-1)\frac{\pi y}{b}]}\right\}
\label{u}
\end{multline}
in which the no slip boundary conditions at the two vertical walls
($u_{x \pm}(x, y=\pm b/2, z)=0$) have been taken into account.
Each Fourier mode, ($(2n-1)\pi/b)$, involves two constants for each
fluid, $a_{\pm n}$ and $b_{\pm n}$.
 These four constants are determined by using the no slip boundary conditions at the bottom and top of the cell ($u_{x +}(x, y, z=0)=u_{x -}(x, y, z=H)=0$) and the continuity of the velocity ($u_{x +}(x, y, z=h)=u_{x -}(x, y, z=h)$)
and of the shear stress
($\eta \, \partial u_{x +}(x, y, z=h)/\partial z =\eta \, \partial u_{x -}(x, y, z=h)/\partial z$) at the interface.
\begin{figure}
\begin{center}
 \includegraphics[width=14cm]{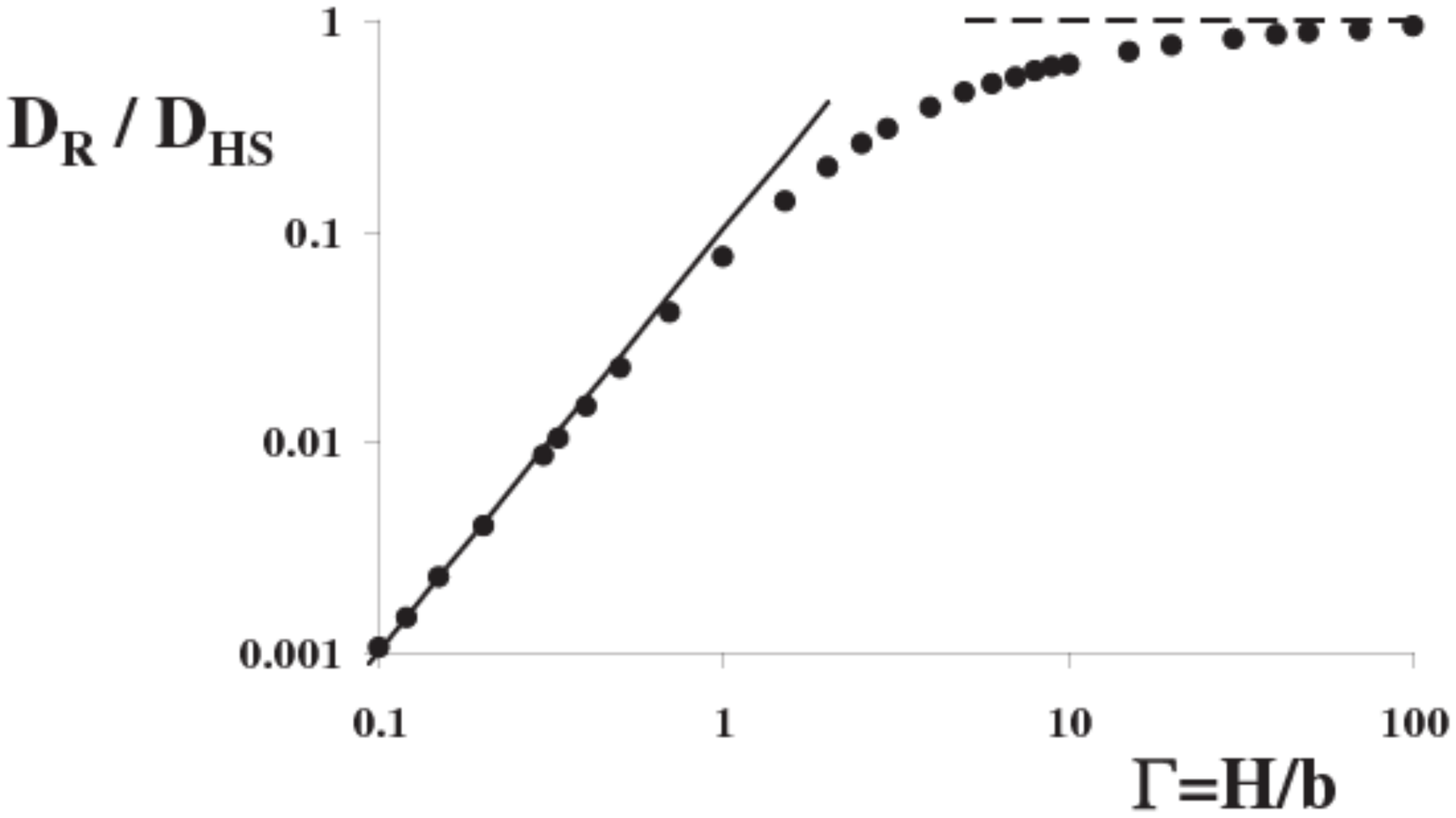}
            \caption{Log-log plot of the normalized lock-exchange diffusion coefficient,
$D_R/D_{HS}$, versus the  aspect ratio $\Gamma=H/b$, for rectangular cells.
$D_R$ is normalized by the diffusion coefficient $D_{HS}=b^2 \,H \Delta\rho g/(12 \, \eta)$
(eq. (\ref{diffdarcy}), obtained for the Hele-Shaw cell limit,
also corresponding to a $2D$ porous medium. Accordingly, the latter regime corresponds to the
dashed straight line ($D_R/D_{HS}=1$) in this representation.
The solid straight line corresponds to the $2D$ Stokes limit (eq. (\ref{D2Dstokes})), leading to a
slope $2$ in this representation.}
\label{Dcrossover}
 \end{center}
\end{figure}
Combining the so-obtained expressions for the velocity with eq. (\ref{pressuregrad})
and the no net flux condition (eq. (\ref{netflux})),
one obtains the horizontal flux of the heavy fluid (eq. (\ref{diffusiveflux})):
\begin{equation}\label{hsflux}
q=-D_{HS} f_\Gamma(h/H)\frac{\partial\,h}{\partial\,x}
\end{equation}
with
\begin{eqnarray}
  & & D_{HS}= \frac{b^2 H \Delta \rho g}{12 \eta}  \label{d2Dbis}\\
  & & f_\Gamma(\psi)=
\frac{\psi+\alpha_\Gamma(\psi)}{1-\gamma_\Gamma}(1 - \psi -
\alpha_\Gamma(\psi)-\gamma_\Gamma) - \delta_\Gamma(\psi) \label{g}
\end{eqnarray}
where
\begin{eqnarray}
\alpha_\Gamma(\psi) & = & \frac{1}{\Gamma}\sum_{n = 1}^{\infty}
   \frac{96\,\left(1 + e^{\Gamma\left(1-\psi \right) \,\left(2\,n-1\right) \,\pi} \right) \,
        \left(1 - e^{\Gamma\;\psi  \,\left(2\,n-1\right) \,\pi} \right) }{
        \left(1 + e^{\Gamma\,\left(2\,n-1\right) \,\pi} \right)
\,{\left(2\,n-1 \right) }^5\,{\pi }^5} \label{A1} \\
\label{B1}
\delta_\Gamma(\psi) & = &  \label{D1}  \frac{1}{\Gamma}\sum_{n =
1}^{\infty}\frac{48\,{\left(1 -
              e^{\Gamma\left(1-\psi \right) \,\left(2\,n-1\right) \,\pi } \right) }^2\,
          {\left(1 - e^{\Gamma\;\psi \,\left(2\,n-1\right) \,\pi } \right) }^2}
{\left(-1+e^{2\,\Gamma\,\left(2\,n-1\right)\,\pi}\right)\,{\left(2\,n-1 \right)}^5\,{\pi }^5} \\
  \gamma_\Gamma &= & \frac{1}{\Gamma}\sum_{n=1}^{\infty}
\frac{192\,\tanh \, (\frac{\Gamma\,\left(2\,n-1 \right) \,\pi }{2})}
         {{\left(2\,n-1 \right) }^5\,{\pi }^5}
\end{eqnarray}
Eq. (\ref{hsflux}) admits a self-similar solution, $h(\zeta)=H\, \psi(\zeta)$,
with the similarity variable $\zeta=x/\sqrt{D_{HS}\, t}$,
which obeys eq. (\ref{similarity}) or eq. (\ref{similaritybis}).
 As previously, it is easier to compute the solution $\zeta (\psi)$ of
eq. (\ref{similaritybis}) subject to the corresponding asymptotics,
$\zeta=-\zeta_0 +8\, \Gamma^2 \, \psi^{3}/(3 \,\zeta_0)$ in the vicinity
of the boundary, $\psi=0$.
We solve this equation using the shooting method previously described
and using Mathematica Software.
The solutions $h(\zeta)$ are plotted in Fig. \ref{GCprofil} for different values of the cell aspect ratio $\Gamma$.
We notice that, in contrast with Darcy predictions (straight line in Fig. \ref{GCprofil}), but similarly to the case of the $2D$ Stokes flow,
the profiles, $h(\zeta)$, exhibit vertical slopes at the edges of the cell.
We note also that such vertical slopes were observed in the experiments
 by \cite{seon07} and \cite{huppert95}.
When comparing their experiments in a Hele-Shaw cell with Darcy predictions,
the latter authors
reported that ''Some discrepancies develop near the leading edge of the current
as a result of the increasing importance of the bottom friction at the nose''
(Fig. 2 of \cite{huppert95}). This mismatch will be addressed in the next section. According to the so-obtained profiles, stationary in the similarity variable $\zeta$, the leading and trailing edges of the front spread as the square root of time, and a lock-exchange diffusion coefficient, dependent on the cell aspect ratio, can be defined:

\begin{equation}\label{DRect}
    D_{R}= \frac{b^2 H \Delta \rho g }{12\eta} \, F(\frac{H}{b})=D_{HS}\, F(\frac{H}{b})
\end{equation}
Fig. \ref{Dcrossover} displays a log-log plot of the normalized
rectangular cell lock-exchange diffusion coefficient, $D_R /D_{HS}=F(H/b)$, versus the
aspect ratio $\Gamma=H/b$.
At small aspect ratios, $\Gamma <1$, the diffusion coefficient falls on top of the full line of slope $2$, which corresponds to the $2D$ Stokes flow between
boundaries distant of $H$ ($b \rightarrow \infty$, eq. (\ref{D2Dstokes})).
At large aspect ratios, $\Gamma \rightarrow 100$,
the diffusion coefficient approaches the
dashed line, $D_R /D_{HS}=1$, obtained for the $2D$ homogeneous
porous medium case (eq. (\ref{diffdarcy})).
We note that the latter case,
which corresponds to a Hele-Shaw cell of infinite aspect ratio,
overestimates the lock-exchange diffusion coefficient, by a relative amount of
about $30\%$, for aspect ratios as large as $\Gamma=10-20$.


\section{$2D$ Stokes-Darcy model for lock-exchange in a rectangular cross-section channel}

The above-mentioned failures of the $2D$ Darcy model at finite aspect ratios may come from the velocity slip condition at
the bottom and top edges of the cell ($z=0$ and $z=H$, respectively).
This non physical condition is indeed required by the use of Darcy equation for the flow, which neglects the momentum diffusion in the presence of velocity gradients, in the plane of the cell ($z-x$ plane).
The momentum diffusion may however be taken into account in
$2D$, through the so-called Stokes-Darcy equation
(see \cite{bizon97, ruyer-quil01, martin02, zeng03}), which is similar to the
Darcy-Brinkman equation used in porous media (see \cite{brinkman47}).
This $2D$ model enables to handle discontinuities such as cell edges, gap
heterogeneities and fluid interfaces
(\cite{ruyer-quil01, martin02, zeng03, talon03}) and was successfully applied in the study of Rayleigh-Taylor instability
(\cite{martin02,fernandez02,graf02}), of dispersion in heterogeneous fractures
(\cite{talon03}) and of chemical reaction fronts (\cite{martin02b}).
Although our present case of interest can be handled with $3D$ Stokes calculations, it is of interest to test the applicability of the $2D$ Stokes-Darcy model to the case of deep and narrow cells.
Indeed, such a $2D$ model, once validated, could be a useful tool to
 address the issue of more complicated cases,
such as gravity currents in the presence of viscosity contrasts, or in fractures with aperture heterogeneities.

\noindent In this model, the flow in the rectangular cell (Fig. \ref{schema})
is assumed to be parallel to the plates
($\vec u(x, y, z)=(u_x(x, y, z), 0, u_z(x, y, z)$) with a Poiseuille parabolic
profile across the gap (the key assumption). Using the
Stokes equation with this $y$ dependency, the gap-averaged
fluid velocity $\vec U(x, z)=\frac{1}{b}\int_{-b/2}^{b/2} \vec u (x, y,  z)dy$,
follows  a Stokes-Darcy (SD) equation which reads here for the
horizontal component of the velocity:
\begin{equation} \label{eq:DB}
-\frac{12 \eta}{b^{2}}U_{x \pm} (x, z) + \beta \;\eta \nabla^{2} U_{x \pm}
(x, z)=\frac{\partial P_\pm}{\partial x}
\end{equation}
The first term on the left hand side of eq. (\ref{eq:DB}) and the pressure gradient correspond to the Darcy's law (eq. (\ref{darcy})) with a permeability
$\kappa=\frac{\displaystyle{b^2}}{\displaystyle{12}}$ for the Hele-Shaw cell as
mentioned above (eq. (\ref{diffdarcy})).
 The second term on the left hand side of eq. (\ref{eq:DB}) is the Brinkman correction to the Darcy equation (see \cite{brinkman47}), which involves an effective viscosity, $\beta \eta$. This effective viscosity may be taken equal to the one of the fluid ($\beta=1$) for the sake of simplicity (or to enable the matching with a $2D$ Stokes regime at $\Gamma \rightarrow 0$).
However, \cite{zeng03} showed that in the Hele-Shaw cell regime (at large  $\Gamma$), the effective viscosity was slightly higher, with $\beta= 12/\pi^2 \simeq 1.215$.
\begin{figure}
\begin{center}
 \includegraphics[width=14cm]{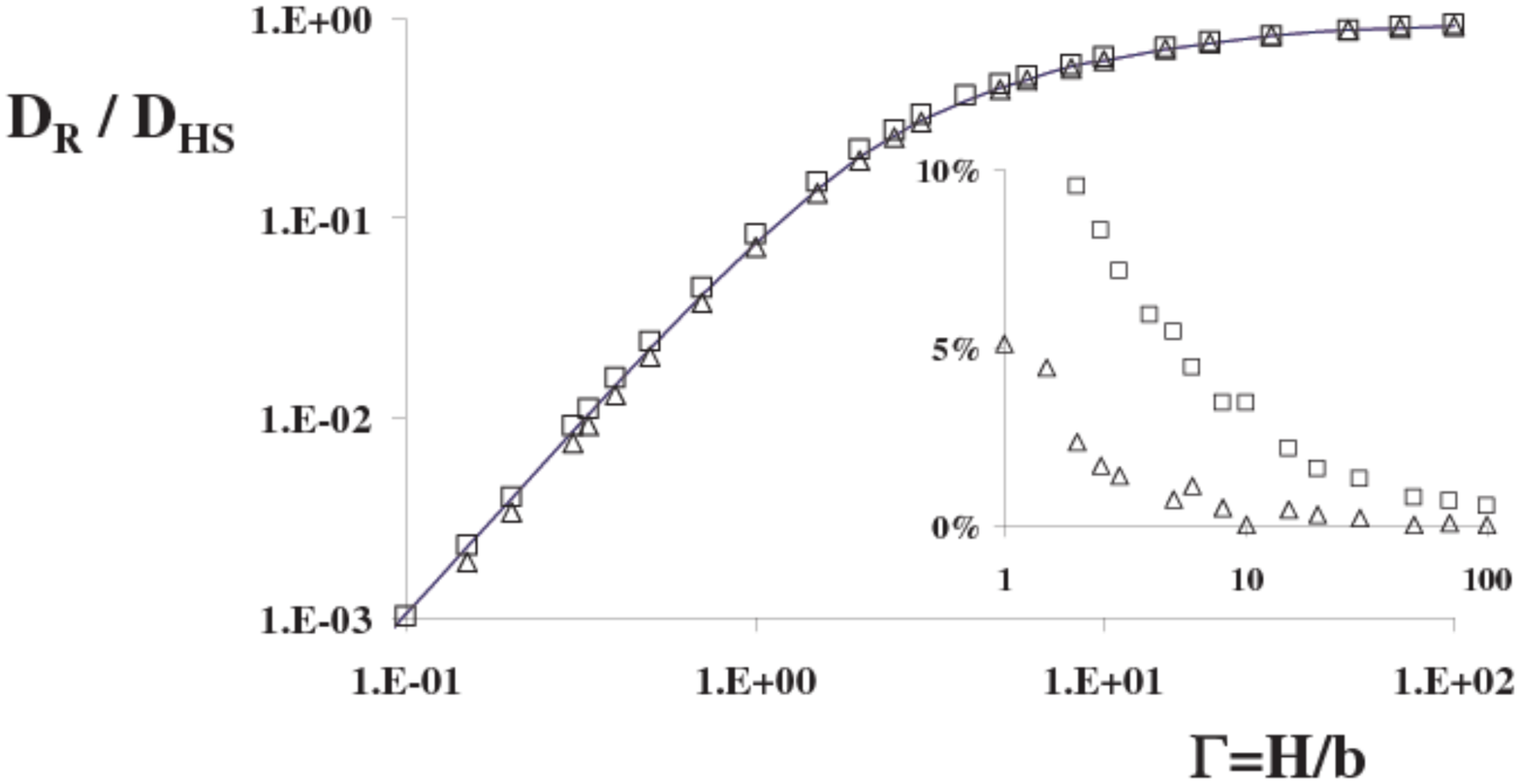}
  \caption{Log-log plot of the normalized lock-exchange diffusion coefficient,
versus the aspect ratio $\Gamma=H/b$, for rectangular cells. The data points,
obtained using the $2D$ Stokes-Darcy model (eq. (\ref{diffSD})) with Brinkman viscosity factors $\beta=1$ (open squares) and $\beta=12/\pi^{2}$ (open triangles) are compared to the full $3D$ results already shown in Fig. \ref{Dcrossover} and depicted here as a solid line. The inset gives the relative difference between the $2D$ Stokes-Darcy model
($\beta=1$, open squares and $\beta=12/\pi^{2}$, open triangles) and the $3D$ calculations. Note that these data were obtained by the difference between values of accuracy of the order of a few
$10^{-3}$, which results in the small dispersion observed in the figure.}
\label{diffSD}
 \end{center}
\end{figure}
At a given location $x$, integrating  eq. (\ref{eq:DB}) leads to the two
velocity  profiles matching the no slip boundary conditions at the bottom and the top boundaries ($u_{x +}(x, 0)=u_{x -}(x, H)=0$) and the continuity of the velocity ($u_{x -}(x, h)=u_{x +}(x, h$)) and of the shear stress
($\beta \,\eta \,\partial u_{x -}(x, h)/\partial z=
\beta \,\eta \,\partial u_{x +}(x, h)/\partial z$) at the interface.
 Using the no net flux condition,
$\int_{0}^{h}u_{x +}dz +\int_{h}^{H}u_{x -}dz=0$, and eq. (\ref{pressuregrad}), we obtain the horizontal flux (eq. (\ref{flux})):
\begin{equation}\label{dbflux}
q(x)=-D_{HS} \, f_{SD}(h/H)\,\frac{\partial\,h}{\partial\,x}
\end{equation}
where $D_{HS}$ was already given in eq. (\ref{diffdarcy}) and the reduced flux function is equal to:


\begin{multline}
    f_{SD}(\psi)  = \frac{1}{4 d ( d -\tanh d )}\left\{ 2 + 4\,d^2\,( 1
- \psi ) \,\psi -  d\,\frac{
 3\,\cosh (2\,d) + \cosh (2\,d\,( 1 - 2\,\psi ) ) }{\sinh (2\,d)}\right. \\
 \left. +4\, d\,\frac{( 1 - \psi ) \,\cosh (2\,d\,( 1 - \psi) ) + \psi\,
 \cosh
(2\,d\,\psi)}{\sinh (2\,d)} \,- 2\,
\frac{\cosh (d\,( 1 - 2\,\psi ) )}{\cosh(d)} \right\}
\label{fluxStokesDarcy}
\end{multline}
where

\begin{equation}\label{d}
d=\sqrt{\frac{H^{2}}{4 \kappa \beta}}=\sqrt{\frac {3  }{ \beta}} \, \Gamma
\end{equation}
and $\kappa=b^{2}/12$. A comparison of the full $3D$ calculations for a rectangular channel of aspect ratio $\Gamma=H/b$ with this $2D$ approximation can be performed on the flux functions, $f_\Gamma (\psi)$ (eq. (\ref{g})) and $f_{SD}(\psi)$ (eq. (\ref{fluxStokesDarcy})).
 These two flux functions are close to each other,
within a few per cents. In order to address the comparison in the range of interest for the Hele-Shaw assumption, i.e. $\Gamma \gg 1$, let us analyze the limit $\Gamma \rightarrow \infty$ ($d \rightarrow \infty$), which gives
 \begin{equation}\label{DBinft}
f_{SD,\Gamma \rightarrow \infty}\simeq \psi (1-\psi)-(\frac{3}{4}-\psi (1-\psi))
\sqrt{\frac {\beta}{3}} \;\frac{b}{H} + O\left((\frac{b}{H})^2\right)
\end{equation}
for the Stokes-Darcy flux and
\begin{equation}\label{3Dinft}
f_{\Gamma,\Gamma \rightarrow \infty}\simeq \psi (1-\psi)-(\frac{3}{4}-\psi
(1-\psi)) \frac{186 \, Zeta(5)}{\pi^{5}} \; \frac{b}{H}  +
O\left((\frac{b}{H})^2\right)
\end{equation}
for the full $3D$ rectangular cell flux (with $Zeta(5)=\sum_{1}^{\infty} n^{-5}=1.03693$, the value of the Riemann-Zeta function). We note that the leading term of both series corresponds to the expected porous media Darcy limit (eq. (\ref{fluxDarcy})) with a permeability $\kappa=b^{2}/12$. However, the next order term ($O(b/H)$) is not the same, unless one chooses for the factor $\beta$,
\begin{equation}\label{eta}
\beta=3 \; (\frac{186 \, Zeta(5)}{\pi^{5}})^{2}\simeq 1.192 \end{equation}
which is very close to the value $12/\pi^{2} \simeq 1.215$ found
by \cite{zeng03} and to the value $6/5$ proposed by \cite{ruyer-quil01}.
The lock-exchange diffusion coefficient has been computed, with the same procedure as above, by integrating eq. (\ref{similaritybis}), using $f_{SD}(\psi)$, from $\psi=0.5$, and matching the asymptotics, $\zeta=-\zeta_0 +8\, d^{2} \, \psi^{3}/(9 \,\zeta_0)$
in the vicinity of the boundary, $\psi=0$.
The so-obtained lock-exchange diffusion coefficients, calculated for two different values of $\beta$ ($\beta=1$ and $\beta=12/\pi^{2}$) are compared to the $3D$ calculations in Fig. \ref{diffSD}.
 The data for both values of $\beta$ are indeed very close to the $3D$ data
over the whole range of aspect ratios, $\Gamma=1-100$.
The inset of Fig. \ref{diffSD} gives the percentage of error for the two values of $\beta$. We note that these data were obtained by the difference between values of accuracy of the order of a few $10^{-3}$, which results in the small dispersion observed in the inset of Fig. \ref{diffSD}.
As expected, the results at large $\Gamma$ (in the Hele-Shaw regime) are closer to the $3D$ full problem for $\beta=12 /\pi^{2}$  than for $\beta=1$.
We point out that, whereas the Brinkman term does bring a significant correction, the exact value of the Brinkman viscosity factor $\beta$ is however not crucial: For instance, for $\Gamma=10$, we obtained a
diffusion coefficient $3.5\%$ smaller than the $3D$ value for $\beta=1$ and
$0.1\%$ larger for $\beta=12/\pi^{2}$, to be compared to the $30\%$ of error if
the cell was assumed to be of infinite aspect ratio (Hele-shaw limit) as in \cite{huppert95}. In conclusion of this comparison, we have shown that the $2D$ Stokes-Darcy model for lock-exchange in a rectangular cell captures quite accurately  the effect of the finiteness of the cross-section aspect ratio. By using the correct $\beta$ value, the error in the model is smaller than $5\%$ for aspect ratios larger than $\Gamma>1$.

\section{Experiments}

In this section, we will present experimental measurements of the diffusion
coefficient in Hele-Shaw cells of different aspect ratios and we will compare
them with our computed values.

\begin{figure}
\begin{center}
 \includegraphics[width=14cm]{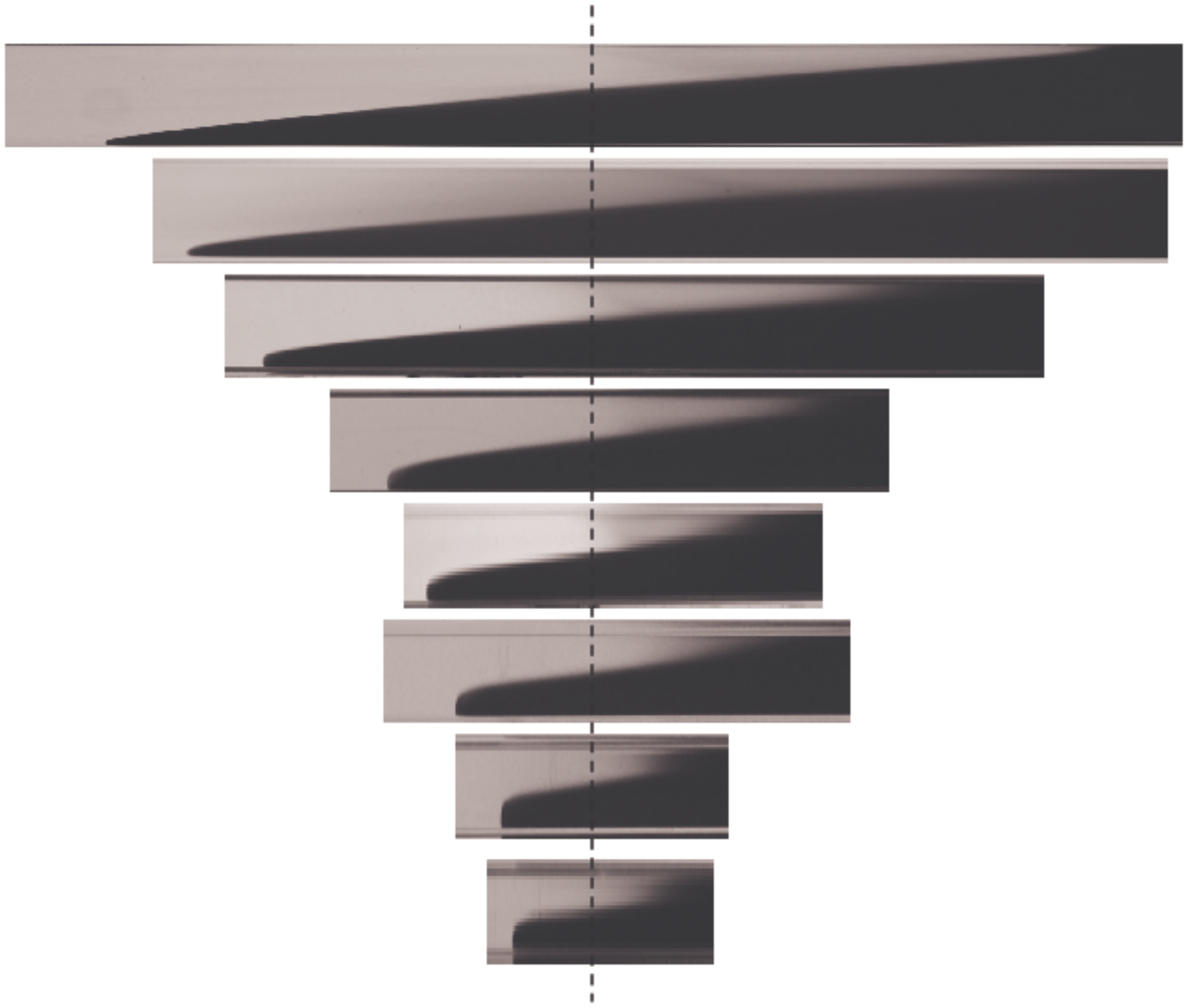}
   \caption{Side view (plane $z-x$) of the lock-exchange interface for rectangular
cells of aspect ratios $\Gamma=H/b=10$, $6$, $2.5$, $1.5$, $1$, $2/3$, $0.4$, $1/3$,
from top to bottom, respectively. The horizontal axis is scaled with $\sqrt{D_{HS}\,t}$,
which allows to observe the decrease of $\zeta_{0}$ as $\Gamma$ decreases.
The vertical dashed line corresponds to $\zeta=0$.}
\label{photos}
 \end{center}
\end{figure}

\noindent We used borosilicate rectangular cells of height $H$ and thickness $b$ and typical length
$30 \,cm$ (Fig. \ref{schema}). The rectangular cross-sections of the cells were
(in $mm^{2}$): $2\times 6$, $2 \times 12$, $2 \times 20$, $3 \times 3$,
$3 \times 9$, $3 \times 30$, $4 \times 6$, $4 \times 10$, $6 \times 6$.
Each cell was used with one side or the other held vertically, leading to two aspect ratios per cell. With such values, we covered a wide range of aspect ratios, from $\Gamma=H/b= 1/10$ to $10$. We used, as Newtonian miscible fluids,
aqueous solutions of natrosol and calcium chlorite.
The fluids had equal viscosities, which were fixed by the polymer concentration and measured with an accuracy of $1\%$. The fluid densities were adjusted by addition of salt and measured with an accuracy of $0.01\%$.
\begin{figure}
\begin{center}
 \includegraphics[width=14cm]{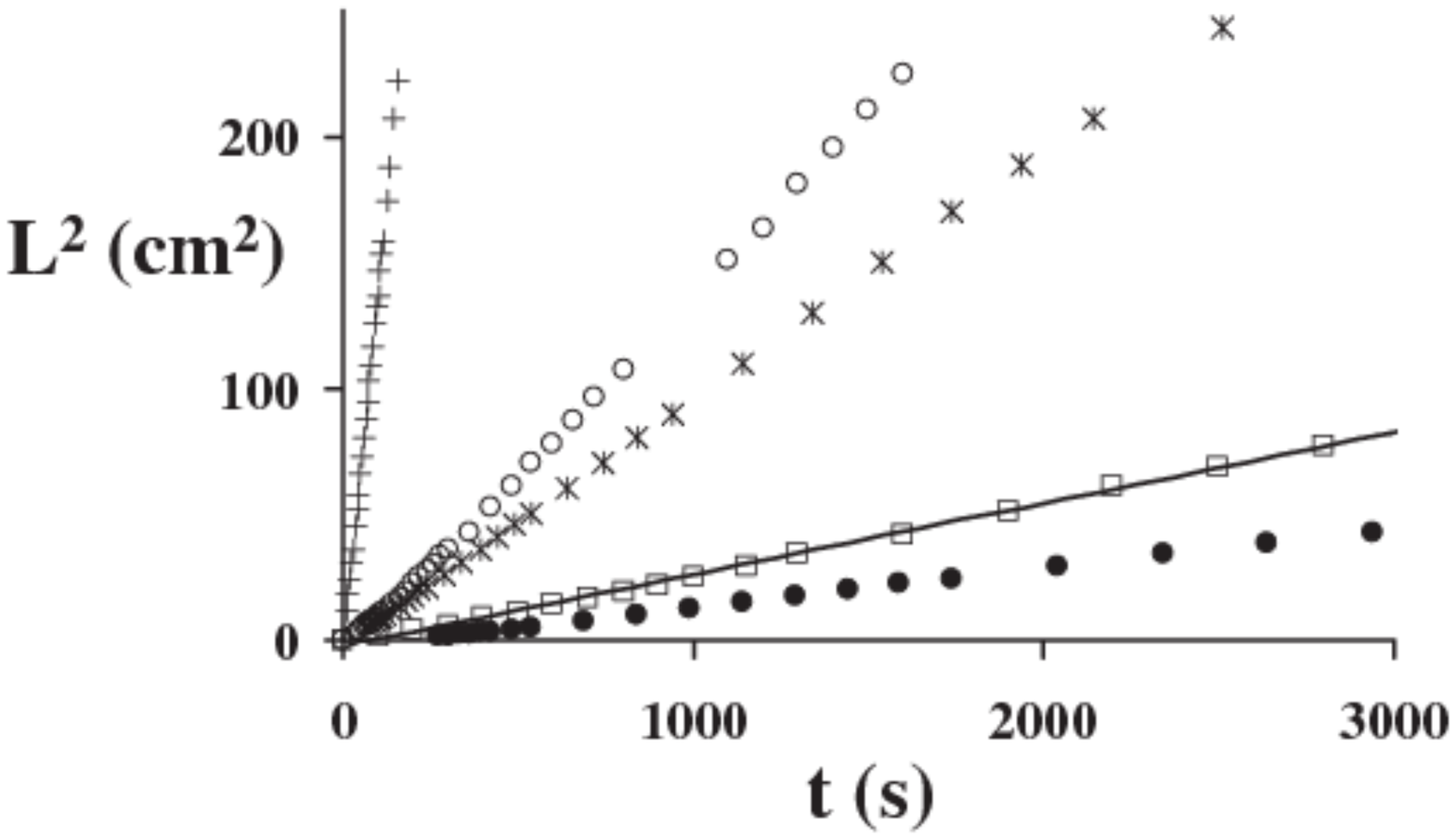}
  \caption{Plot of the square of the spreading distance of the leading edge of the front
versus time for five cells: ($\bullet$) $H=3\,mm$, $b=30 \,mm$;($\square$)
$H=6\,mm$, $b=2 \,mm$; ($\ast$) $H=6\,mm$, $b=4 \,mm$; ($\circ$)
$H=b=6 \,mm$; ($+$) $H=30\,mm$, $b=3 \,mm$.
 The solid line is a linear fit to the data, the slope of which gives the
diffusion coefficient plotted in Fig. \ref{diffmanip} (top).}\label{xsquaremanip}
 \end{center}
\end{figure}
The overall accuracy in $D_{HS}$ was typically $5\%$, when taking into account
the above accuracies in viscosities and densities and the inherent temperature
variations during the experiments. The viscosities and the densities of the fluids were chosen to satisfy two experimental requirements. The experiments must be fast enough in order to prevent any significant molecular mixing of the fluids and one should be able to put the two fluids in contact without mixing. The latter condition requires a rather large density contrast and large viscosities. With our cell sizes, a good compromise  was obtained with a density contrast of about $1 \%$ and typical viscosities in the range, $10-50 \, mPa.s$, leading to a lock-exchange diffusion coefficient ranging from $10^{-3} cm^{2}/s$ to $1 cm^{2}/s$. The typical Reynolds number,
built with the gap of the cell of these experiments is smaller than $0.1$.
For each experiment, the cell was first held with its axis $Ox$ vertical.
The fluids were successively slowly injected, with the lighter fluid on top of the heavier. Then the cell was closed and put in the desired position, with its axis $Oz$ vertical in a few seconds. The development of the lock-exchange pseudo-interface was then recorded thanks to a video camera.
\begin{figure}
\begin{center}
 \includegraphics[width=14cm]{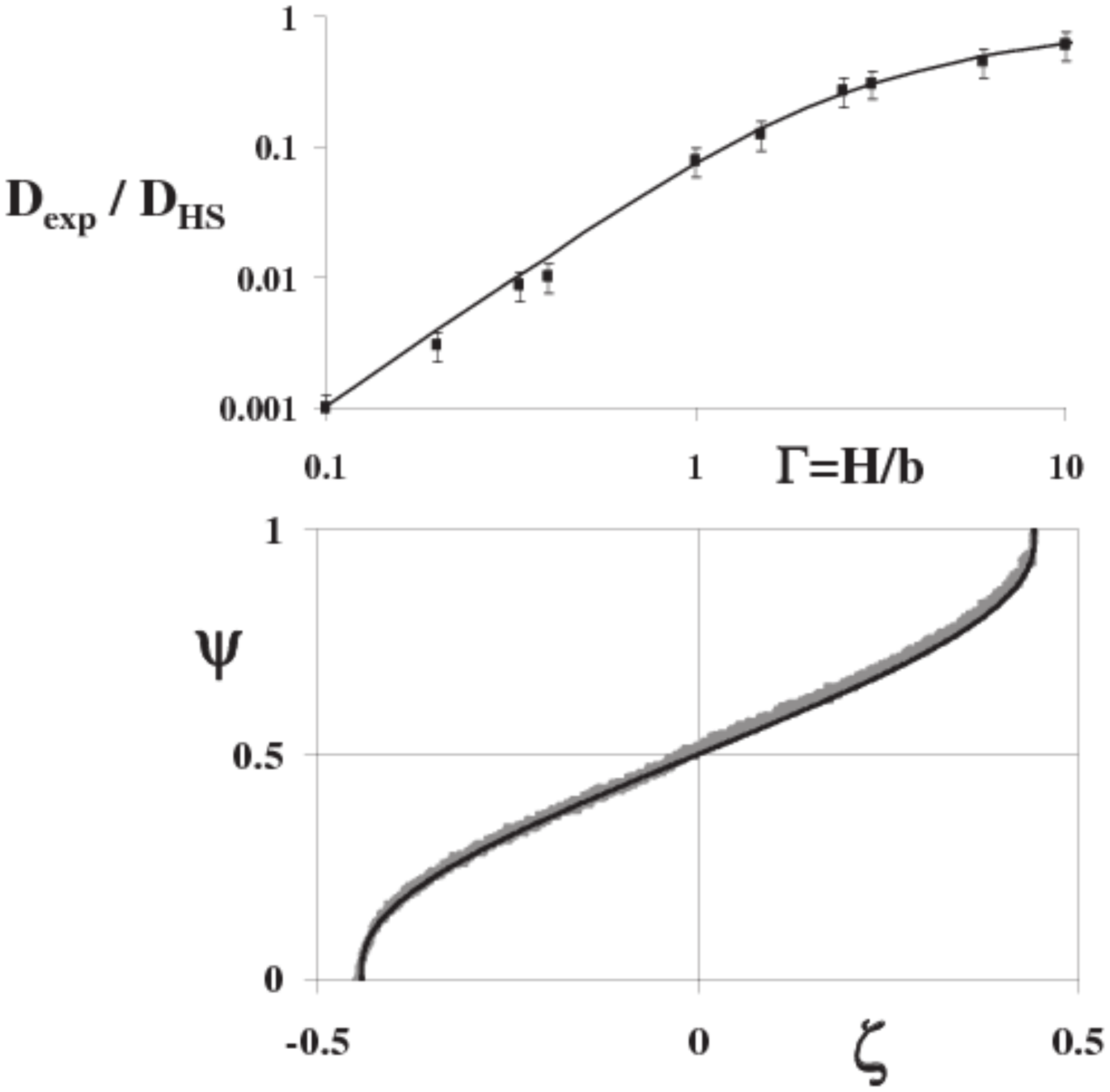}
            \caption{Top: Log-log plot of the normalized measured lock-exchange diffusion coefficient, $D_{exp}/D_{HS}$ (open squares), versus the aspect ratio, $\Gamma=H/b$, of the cell.
            The solid line corresponds to the $3D$ results (same as in Fig. \ref{diffSD}).
             Bottom : Superimposition of the measured pseudo-interface between the two fluids (grey fuzzy line) and of the theoretical profile (full dark line) corresponding to Fig. \ref{GCprofil}. The aspect ratio is $\Gamma=4$.}
            \label{diffmanip}
 \end{center}
\end{figure}
Typical pictures (side view in the plane $z-x$) are given in Fig. \ref{photos}
for cells of different aspect ratios. The horizontal axis is scaled with $\sqrt{D_{HS} \, t}$, so that one can see the decrease of $\zeta_{0}$ as $\Gamma$ decreases. With this representation using the self-similar variable $\zeta=x/\sqrt{D_{HS} \,t}$, the  profiles are stationary. One may notice that the trailing edge is fuzzy. This can be attributed to the stick condition at the upper wall: The dark dense fluid does stay at
the walls for a long time, in particular in the corners of the cross-section.
The same phenomenon takes place at the bottom of the cell, but the presence of transparent light fluid has little effect on the turbidity of the heavy dark fluid, and is therefore not noticeable on the pictures.
It is worth noting that the shape of the leading edge evolves from an edge at large aspect ratios $\Gamma$ to a more and more step-like shape as $\Gamma$ decreases. It should be noticed that for small aspect ratios,
although it is rather difficult to take pictures, a top view of the cell reveals a mild spanwise dependency of the interface, but we do not observe the
spanwise lobe-and-cleft instability reported by \cite{simpson72}.
  For each experiment, the locations of the leading and trailing edges of the front were measured in time. Fig. \ref{xsquaremanip} gives the variations of the square of the spreading distance versus time for five cells. It is worth noting that the dependency is almost linear: Therefore a linear fit provided
the lock-exchange diffusion coefficient, with a typical accuracy of $20\%$.
Fig. \ref{diffmanip} (top) displays the so-obtained normalized lock-exchange diffusion coefficient as a function of $\Gamma$. One can see
that the agreement with the $3D$ calculations over the two decades of our measurements is rather good. We note that for the large aspect ratio limit of the experiments (up to $\Gamma=10$), the Hele-Shaw cell limit is not reached, and would underestimate, by $30\%$, the lock-exchange diffusion coefficient.
This result thus confirms that for such aspect ratios, one should either compute the full $3D$ Stokes equation or use the Stokes-Darcy model to obtain the correct behaviour. We also note that our calculation still holds for aspect ratios as small as $\Gamma = 0.1$. This result is quite unexpected since for such aspect ratios, some spanwise dependency of the profile was observed, and the hypothesis of the interface surface, $h(x, y, z)$, invariant in the $y$
direction is certainly broken.
The bottom of Fig. \ref{diffmanip} displays the superimposition of
the theoretical and the experimental interfaces between the fluids, for
 an aspect ratio $\Gamma=4$. The agreement between the two is rather good, thus
 validating our model. Such an agreement is rather surprising as our small slope assumption is violated at the edges of the gravity current. This agreement, already emphasized by \cite{huppert82} and \cite{seon07}, is likely to be common to viscosity dominated gravity current without surface tension.

\section{Conclusion}
The viscous lock-exchange diffusion coefficient
reflects the competition between the buoyancy driving effect and the viscous damping, and depends on the geometry of the channel. We give the backbone to calculate this coefficient in different configurations: We recall its computation for a porous medium already found by \cite{huppert95}, and compute it for a $2D$ Stokes flow between
two parallel horizontal boundaries separated by a vertical height, $H$.
This result is in agreement with \cite{hinch07} (unpublished)  and in reasonable agreement with recent computations  by \cite{taghavi09}. Using a quasi-parallel flow assumption, we have calculated the pseudo-interface profile between the two fluids and the diffusion coefficient of
viscous lock-exchange gravity currents for a rectangular channel
(horizontal thickness $b$, vertical height $H$) of any aspect ratio ($H/b$).
This analysis provides a cross-over between the $2D$ Stokes flow between
two parallel horizontal boundaries separated by a vertical height, $H$,
 and the Hele-Shaw cell limit (applying for $H/b>100$).
 Moreover, the shape of our profiles allows
to account for the discrepancy observed at the nose of the gravity current in
the experiments by \cite{huppert95}. The agreement, obtained despite the failure of the lubrication assumption at the edges of the current, should deserve however further theoretical investigation. Our calculations  of the diffusion coefficient and of the shape of the profile have also been convincingly compared to new experiments carried out in cells of various aspect ratios ($1/10-10$).
We have also calculated the lock-exchange diffusion coefficient
for the same rectangular cells, using the $2D$ Stokes-Darcy model.
This model is shown to apply to aspect ratios $H/b>1$, provided
that the appropriate Brinkman correction is used. Such a $2D$ model may be useful to describe gravity currents
with a finite volume of release, with fluids of different viscosities, or in
 heterogeneous vertical fractures.

\section{Acknowledgement}
{\it This work was partly supported by CNES (No 793/CNES/00/8368),
ESA (No AO-99-083), by R\'{e}seaux de Th\'{e}matiques de Recherches Avanc\'{e}es
''Triangle de la physique'', by the Initial Training Network (ITN) ''Multiflow'' and
 by French Research National Agency (ANR) through the
''Captage et Stockage du CO$_{\mbox 2}$'' program (projet CO-LINER No ANR-08-PCO2-XXX).
All these sources of support are gratefully acknowledged.}

\end{document}